\documentclass[amsmath,amssymb,amsfonts,aps,pre,superscriptaddress,bibnotes,showpacs,showkeys,longbibliography,10pt,nobibnotes]{revtex4-2}
\usepackage{graphicx}
\usepackage{caption,subcaption}
\usepackage{amsmath,amssymb,amsfonts}
\usepackage{sidecap}
\usepackage{xcolor}
\usepackage{lipsum}
\usepackage{physics}
\usepackage{hyperref}

\newcommand{\me}{\mathrm{e}}
\DeclareMathOperator\erfi{erfi}
\newcommand{\kb}{k_{\rm B}}
\newcommand{\Teff}{T_{\rm eff}}

\begin{document}

\graphicspath{ {Images/} }

\title{Dynamic control of the Bose-Einstein-like condensation transition in scalar active matter}

\author{Jonas Berx}
\email{jonas.berx@nbi.ku.dk}
\affiliation{Niels Bohr International Academy, Niels Bohr Institute, University of Copenhagen, Blegdamsvej 17, 2100 Copenhagen, Denmark}
\date{\today}

\begin{abstract}
The dynamics of a generic class of scalar active matter exhibiting a diffusivity edge is studied in a confining potential where the amplitude is governed by a time-dependent protocol. For such non-equilibrium systems, the diffusion coefficient vanishes when the single-particle density field reaches a critical threshold, inducing a condensation transition that is formally akin to Bose-Einstein condensation. We show that this transition arises even for systems that do not reach a steady state, leading to condensation in finite time. Since the transition can be induced for a fixed effective temperature by evolving the system, we effectively show that the temporal coordinate constitutes an alternative control parameter to tune the transition characteristics. For a constant-amplitude protocol, our generalised thermodynamics reduces in the steady-state limit to earlier results. Lastly, we show numerically that for periodic modulation of the potential amplitude, the condensation transition is reentrant.
\end{abstract}

\maketitle

\section{Introduction \label{sec:intro}}
The exploration of self-organising systems through intrinsic microscopic activity has led to the discovery of remarkable phenomena within the realm of active matter \cite{Gompper2020}. Such active non-equilibrium systems break detailed balance at a local level and can lead to collective effects such as motility-induced phase separation (MIPS), where no attractive interactions are present to induce phase separation \cite{Bechinger2016,Cates2015}.

The key to understanding the rich phenomenology of active systems lies in the formulation of minimal models that capture the essential physics \cite{Buttinoni2013,Gompper2020}. Recently, a class of such active systems, distinguished by a diffusivity edge inducing condensation at steady states, has emerged as a generic mean field description for density-dependent diffusivity, in which some mean field dynamical features are represented \cite{Golestanian_2019,Mahault2020,Meng2021,Berx_2023}. This diffusivity edge, characterised by a critical density threshold beyond which particle mobility vanishes, introduces a unique and intriguing facet into the dynamics of scalar active matter.

The concept of a diffusivity edge was introduced phenomenologically in Ref.~\cite{Golestanian_2019}, where confining the active system in a harmonic potential led to a condensation transition at the ground state of the potential, where particles aggregate, exhibiting striking similarities with Bose-Einstein condensation \cite{London1938,Ziff1977}. This BEC-like condensation transition was shown through numerical simulation to occur in a model of magnetic microswimmers confined in a quasi-one-dimensional channel \cite{Meng2021}. The notion of a diffusivity edge was subsequently extended to periodic potentials in arbitrary dimensions \cite{Mahault2020}, where co-existing point-like condensates formed at all potential minima. The shallowness of the potentials with respect to the effective thermal energy then quantitatively affects the transition, leading to a non-universality in the values of the scaling exponents.

Introducing an external driving force into the periodically confined system was shown to lead to qualitatively distinct stationary regimes depending on the amplitude of the driving force with respect to the potential barriers \cite{Berx_2023}. For small external forces, the condensation transition was shown to be similar to the transition induced by harmonic confinement, while large forces lead to a reentrant transition, depending on the average density, where the condensation transition is accompanied by a low-temperature evaporation transition. The reentrant regime is characterised by a spatially extended condensate, in contrast with the point-like condensates found for harmonic confinement.

Motivated by the plethora of steady-state phenomena observed in the aforementioned systems, we now shift focus to the dynamical aspects of active matter systems with a diffusivity edge. The paper is structured as follows. In section~\ref{sec:diffusivity}, we introduce our time-dependent harmonically confined active system, and in section \ref{sec:condensation} we characterise the associated condensation transition for two protocol choices. Section~\ref{sec:thermo} then develops a fully time-dependent generalised thermodynamic framework, showing the similarities and differences with Bose-Einstein condensation. Subsequently, in section~\ref{sec:periodic}, we briefly comment on periodic protocols and give a numerical example. In section~\ref{sec:conclusions}, we conclude and give an outlook on future studies.

\section{Scalar active matter with diffusivity edge}\label{sec:diffusivity}
We consider a mean-field description of non-equilibrium matter that is described by an effective single-particle scalar density field $\rho({\bf r},t)$ in $d$ dimensions. The dynamics conserves the number of particles $N = \int\mathrm{d}^d{\bf r}\, \rho({\bf r},t)$ for all times. It satisfies the conservation law $\partial_t \rho + \div\vb{J} = 0$, where the flux is defined as ${\bf J} = -D(\rho)\grad\rho + \rho \vb{v}$. The drift velocity $\vb{v}$ is a result of the external potential $U(\vb{r},t)$, since $\vb{v} = M(\rho) (-{\bf \grad} U)$, and the diffusion coefficient $D(\rho)$ and mobility $M(\rho)$ are generally functions of the density. In the single-particle limit, their ratio defines a tuning parameter $\Teff$ that is equivalent to an effective temperature \cite{Golestanian_2019}, and which is a measure for the density fluctuations at the particle level,
\begin{equation}
    \label{eq:temperature}
    \kb \Teff \equiv \lim_{\rho\rightarrow 0}{\frac{D(\rho)}{M(\rho)}}\,.
\end{equation}

For non-equilibrium systems the fluctuation-dissipation theorem (FDT) is broken, which we implement here by requiring that the single-particle limit of the system is different from the finite-density regime, i.e., we set
\begin{equation}
    \label{eq:FDT_breaking}
    \frac{D(\rho)}{M(\rho)} \neq \lim_{\rho\rightarrow 0}{\frac{D(\rho)}{M(\rho)}}\,.
\end{equation}
Specifically, for large densities we assume the existence of a step function diffusivity edge in the system, i.e., $D(\rho) = M_s \kb \Teff \Theta (\rho_c-\rho)$, where $\rho_c$ is the critical density above which the diffusivity becomes identically zero. Other choices for the diffusivity profile involving, e.g., inhibition or activation are also possible, see, e.g., Ref.~\cite{Golestanian_2019}. Note that one can, in fact, also break the FDT by introducing a mobility edge instead of a diffusivity edge. However, in this work, we will only consider the latter, and for simplicity, we will henceforth fix $M_s = 1$.

We consider the full dynamics of the system by solving the conservation equation, with the spherically symmetric harmonic potential $U(\vb{r},t) = \gamma(t) \vb{r}^2 + \beta(t) \vb{r}$, which allows for the individual tuning of both the stiffness and the position of the ground state by means of the functions $\gamma(t)$ and $\beta(t)$, respectively. The conservation equation is supplemented with the symmetric $d-$dimensional Gaussian initial condition with standard deviation $\sigma$,
\begin{equation}
    \label{eq:initial_condition}
    \psi(\vb{r}) = N \frac{\me^{-\frac{\vb{r}^2}{2\sigma^2}}}{(2\pi\sigma^2)^\frac{d}{2}}\,.
\end{equation}
The general smooth solution remains Gaussian for the non-condensed regime, with a time-dependent variance and amplitude, i.e., $\rho(\vb{r},t) = \rho_0(t) \exp{-\frac{\left(\vb{r} - g(t)\right)^2}{2f(t,\Teff)}}$. The function $f(t,\Teff)$, representing a time-dependent variance, is given by
\begin{equation}
    \label{eq:f}
    f(t,\Teff) = \sigma^2 \me^{-4H(t,0)} + 2\kb \Teff\int_0^t \mathrm{d}\tau\, \me^{-4 H(t,\tau)}\,,
\end{equation}
with $H(t,s) = \int_s^t \mathrm{d}\tau\,\gamma(\tau)$. For $t=0$, it becomes equal to $f(0,\Teff) = \sigma^2$, as required. The function $g(t)$ is given by $g(t) = M_s \int_0^t \mathrm{d}s\, \beta(s) \me^{-2 M_s H(t,s)}$ and represents the time-dependent position of the maximum density. Details of the derivation are given in appendix~\ref{app:density}. Since $\beta(t)$ can be removed from the solution by shifting to a moving frame with time-dependent translation $\vb{r}\rightarrow \vb{r} + M_s \int_0^t \mathrm{d}s\, \beta(s) \me^{-2 M_s H(t,s)}$, we can effectively set $\beta(t) = 0$ without loss of generality.

When the critical threshold density $\rho_c$ is reached, the system develops a singularity at the potential ground state at $r=0$, since the normalisation condition cannot be fulfilled. Hence, the complete density profile is given by
\begin{equation}
    \label{eq:density_total}
    \rho({\bf r},t) = \begin{cases}
        \rho_0(t) \me^{-\frac{r^2}{2f(t,\Teff)}} & \rho_0(t) < \rho_c\,, \\ 
        N_c \delta(r) + \rho_c \me^{-\frac{r^2}{2f(t,\Teff)}} & \rho_0(t) \geq \rho_c\,,
    \end{cases}
\end{equation}
where the ground state density is given by $\rho_0(t) = N/(2\pi f(t,\Teff))^\frac{d}{2}$. At condensation, the ground state density reaches the threshold $\rho_0(T_c) = \rho_c$, which corresponds to $f(t,\Teff) = f(t,T_c)$. 

At this point, we want to stress that condensation can be achieved either by tuning the effective temperature $\Teff$, similar to previous research \cite{Golestanian_2019,Mahault2020,Berx_2023}, or by evolving the system in time, with a suitable (time-dependent) potential. Hence, there exists both a critical temperature $T_c$ and a critical time $t_c$ to condensation.

In this work, we will consider power-law protocols, i.e., $\gamma(t) = \alpha t^\beta$, specifically with the choices $\beta = 0$ and $\beta =1$, which we will refer to as the constant or the linear protocol, respectively, and denote them wherever necessary with the corresponding $\beta$ superscripts. All calculations that follow are straightforward to extend to other choices of $\gamma(t)$. For the choice $\beta = 0,1$, the system is always localised at the ground state. By tuning the parameters $\alpha$ and $\beta$, one can access different regimes of anomalous diffusion \cite{Lillo2000}, which offers a means to represent, e.g., systems with memory \cite{Paraan2006}.

\begin{figure}[htp]
    \centering
    \includegraphics[width=0.75\linewidth]{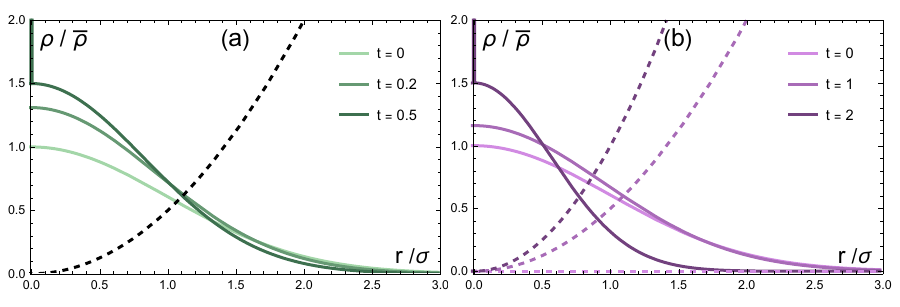}
    \caption{Normalised density profiles (full lines) at different times for {\bf (a)} the constant and {\bf (b)} linear protocols, at fixed $\Teff = 0.5$. Dashed lines show the potential $U(x,t) = \gamma(t) r^2$, where the colours correspond to the ones used for the density profiles of the linear protocol. Thick vertical lines indicate the point-like condensate at $U = 0$}
    \label{fig:density}
\end{figure}

In Fig.~\ref{fig:density}, a projection of the normalised density profiles $\rho/\bar{\rho}$ on a one-dimensional subspace is shown for increasing times, as a function of the coordinate $r$ in the positive real half space. Here, $\bar{\rho} \equiv \rho_0 (t=0) = N/\lambda^d =  N (2\pi\sigma^2)^{-d/2}$ is the initial occupation of the ground state $U = 0$, corresponding to $r = 0$, with $\lambda = \sqrt{2\pi \sigma^2}$ the average initial system size. When the condensation threshold at $\rho = \rho_c$ is reached, a singularity forms at the ground state, shown by a thicker vertical line. When not stated otherwise, we will fix the parameters $\alpha = 1/2$, $\sigma = 1$, $\bar{\rho}/\rho_c = 2/3$, and we work in three spatial dimensions, i.e., $d =3$.

\section{The condensation transition}\label{sec:condensation}

In general, for any $\gamma(t)$, the ratio $\bar{\rho}/\rho_c$ is a fixed quantity and can be used to find the $(\Teff,t)$ phase diagram for the condensation transition, since $\rho_0(t_c,T_c) = \rho_c$ at condensation. From the expressions for $f(t,\Teff)$, i.e., eq.~\eqref{eq:f}, and $\rho_0(t,\Teff)$, we find after some calculation the following relation between the critical temperature $T_c$ and time to condensation $t_c$, for a fixed initial density ratio:
\begin{equation}
\label{eq:critical_Temp_tc}
    k_B T_c = \frac{\sigma^2}{2} \frac{\left[(\bar{\rho}/\rho_c)^\frac{2}{d}\, \me^{4 H(t_c,0)} -1\right]}{\mathcal{G}(t_c)}\,,
\end{equation}
with $\mathcal{G}(t) = \int_0^t \mathrm{d}s\, \exp{4 H(s,0)}$.

\begin{figure}[htp]
    \centering
    \includegraphics[width=0.75\linewidth]{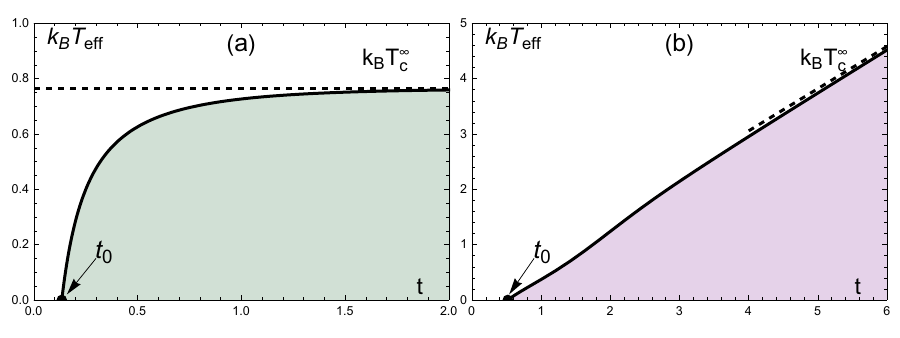}
    \caption{Temperature-time phase diagrams for the condensation transition for {\bf (a)} the constant and {\bf (b)} the linear protocols. The condensed phase is represented by the shaded regions. For the constant protocol, the steady-state transition temperature is given by $\Teff = T_c^\infty$, while for the linear protocol it does not exist, but the transition line is asymptotically linear, $T_c \sim t\, T_c^\infty$ for long times. The zero-temperature transition times $t_0$ are indicated by arrows.}
    \label{fig:constant_phase_space}
\end{figure}

For both protocols, the phase diagram is shown in Fig.~\ref{fig:constant_phase_space} for a fixed initial density ratio $\bar{\rho}/\rho_c = 2/3$, and the critical temperatures are given respectively by
\begin{equation}
    \label{eq:critical_temperatures}
    T_c^{(0)}(t) = \frac{2 \alpha \sigma^2}{\kb}\left(\frac{(\bar{\rho}/\rho_c)^\frac{2}{d} -\me^{-4 \alpha t}}{1-\me^{-4 \alpha t}}\right)\,, \qquad T_c^{(1)}(t) = \sqrt{\frac{2\alpha}{\pi}}\frac{\sigma^2}{\kb} \left(\frac{(\bar{\rho}/\rho_c)^\frac{2}{d} -\me^{-2 \alpha t^2}}{\me^{-2\alpha t^2} \erfi{(\sqrt{2\alpha }t)}}\right)\,.
\end{equation}

For long times, the system confined by the potential with the constant protocol reaches a steady state, and it is only possible to reach condensation in finite time at a temperature strictly below $ T_c^\infty = (\alpha/\kb \pi) (N/\rho_c)^{2/d} $. If we choose $\alpha = k/2$, with $k$ the stiffness of the harmonic trap, the critical temperature in the steady state becomes equal to the previously derived value $ T_c^\infty = (k/2 \pi \kb) (N/\rho_c)^{2/d}$ \cite{Golestanian_2019}. For a linear protocol, however, the critical temperature does not saturate to a constant value in the steady state, but is asymptotically linear in time, with slope $T_c^\infty$. It can be shown that for a general choice of the protocol power $\beta$, the $\beta'$th derivative with respect to time of the asymptotic critical temperature yields $T_c^\infty$ in the high-temperature limit. Note that choosing $\bar{\rho}=\rho_c$ for the constant protocol, i.e., initialising the system in the condensed state, equation~\eqref{eq:critical_Temp_tc} becomes $\kb T_c = 2\alpha\sigma^2$, independent of time.

Since for the constant protocol $T_c(t)$ is an increasing function of time that converges at $T_c^\infty$, the critical temperature is always bounded from above by $T_c^\infty$, and vanishes at a particular time $t_0$ at $\Teff=0$, which is the lower bound for the critical time to condensation; even at zero effective temperature there is a finite amount of time that needs to pass before the density is high enough to induce a condensation transition. This time can be calculated by setting the right-hand-side of eq.~\eqref{eq:critical_Temp_tc} to zero at $t_c = t_0$ and solving for $t_0$. Note that to continue, we will require that $t_0 > 0$, i.e., the system is always initialised in the non-condensed phase. This condition corresponds to the requirement that $\rho_0(t=0) < \rho_c$, such that initial conditions need to be chosen carefully when preparing the system.

The denominator is nonzero for every $t_0 > 0$, so we can eliminate it and solve for $t_0$ in the numerator only, which yields the following simple equation
\begin{equation}
\label{eq:critical_tc}
    H(t_0,0) = -\frac{1}{2d}\ln{\left(\frac{\bar{\rho}}{\rho_c}\right)}\,.
\end{equation}
Since this expression for $t_0$ and equation \eqref{eq:critical_Temp_tc} are general for every choice of $\gamma(t)$ (under the condition that the integral that constitutes $H(t,0)$ converges), the time-dependence of the condensation transition is a universal property of systems exhibiting a diffusivity edge. If the protocol $\gamma(t)$, the initial average density $\bar{\rho}$ and condensation threshold $\rho_c$ are known, the time to condensation at a fixed effective temperature can be tuned exactly. 

The normalisation condition is given by the spatial integral of eq.~\ref{eq:density_total}, and in the condensed regime, this becomes
\begin{equation}
    N = N_c + N\frac{\rho_c}{\rho_0(t)}\,,
\end{equation}
such that the condensate fraction $\phi_c = N_c/N$ is given by
\begin{equation}
    \label{eq:Nc_frac}
    \phi_c(t,\Teff) = 1-\frac{\rho_c}{\rho_0(t,\Teff)} = 1-\left(\frac{f(t,\Teff)}{f(t,T_c)}\right)^\frac{d}{2}\,,
\end{equation}
which is remarkably similar to the equivalent expression for the condensation transition of the ideal Bose gas \cite{Ziff1977}. The function $f(t,\Teff)$ given by
\begin{equation}
    \label{eq:f_constant}
    f^{(0)}(t,\Teff) = \sigma ^2 \me^{-4 \alpha t}+\frac{\kb \Teff \left(1-\me^{-4 \alpha t}\right)}{2 \alpha}\,,
\end{equation}
for the constant protocol, and by 
\begin{equation}
    \label{eq:f_linear}
    f^{(1)}(t,\Teff) = \frac{e^{-2 \alpha t^2} \left(\sqrt{2 \pi \alpha}\, \kb \Teff\erfi{\left(\sqrt{2\alpha} t\right)}+2 \alpha \sigma ^2\right)}{2 \alpha}\,,
\end{equation}
for the linear protocol.

\begin{figure}
    \centering
    \includegraphics[width=0.75\linewidth]{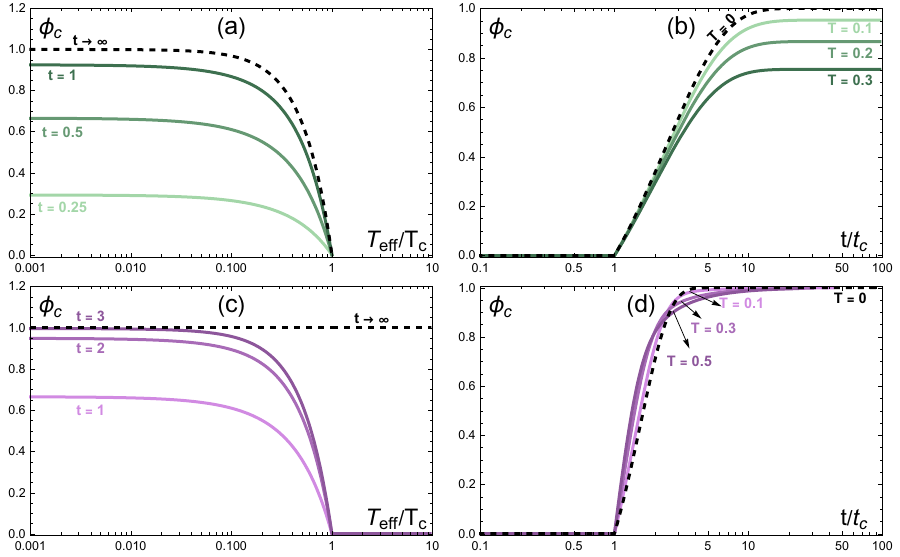}
    \caption{Condensate fraction $\phi_c$ for the constant (a-b) and linear (c-d) protocols, as a function of the effective temperature $\Teff$ (a,c) and time $t$ (b,d). The dashed lines indicate limiting behaviour; for (a,c) it shows the steady-state limit, while for (b,d) it shows the system at $\Teff = 0$.}
    \label{fig:phi_grid}
\end{figure}

The condensate fraction for both the constant and linear protocols converges to $\phi_c = 1-(\Teff/T_c)^\frac{d}{2}$ with $\Teff < T_c$ for long times. In Fig.~\ref{fig:phi_grid}, we show $\phi_c$ as a function of temperature $\Teff$ and time $t$. From panels (b,d), it can be seen that for $\Teff > 0$, the condensate fraction converges to a value $\phi_c^\infty = 1-(\Teff/T_c^\infty)\frac{d}{2}$, for the constant protocol. For the linear protocol, however, the number fraction always converges to unity in the long-time limit. Due to the increasing potential stiffness, particles in the smooth regions are effectively `pushed' into the condensate at $r = 0$ at later times, experiencing an increased effective driving force towards the ground state.

We now turn to the development of a time-dependent generalised thermodynamics for the system.

\section{Generalised thermodynamics}\label{sec:thermo}

The average potential energy $\langle U(t,\Teff)\rangle$ is given by the expectation value of the potential, i.e., 
\begin{equation}
    \label{eq:internal_energy}
    \langle U(t,\Teff)\rangle = \int \mathrm{d}^d{\bf r}\, U({\bf r},t) \rho({\bf r},t) = d
N \gamma(t) f(t,\Teff)\begin{cases}
        1 & \Teff > T_c\,, \\
        \left(\frac{f(t,\Teff)}{f(t,T_c)}\right)^\frac{d}{2} & \Teff\leq T_c\,.
    \end{cases}
\end{equation}

In the long-time limit, the energy converges to
\begin{equation}
    \label{eq:internal_energy_steadystate}
    \lim_{t\rightarrow\infty}{\langle U(t,\Teff) \rangle } = \frac{d}{2} N\kb \Teff\begin{cases}
        1 & \Teff > T_c\,, \\ 
        \left(\frac{\Teff}{T_c^\infty}\right)^\frac{d}{2} & \Teff\leq T_c\,,
    \end{cases}
\end{equation}
where one needs to take into account that the critical temperature is not necessarily finite in this limit, as is the case for, e.g., the linear protocol, where $T_c^\infty$ and consequently also the energy diverge. For short times, however, the energy is a polynomial of the time $t$, with the same exponent as the protocol $\gamma(t)$, i.e., the energy is asymptotically
\begin{equation}
    \label{eq:internal_energy_asymptotic_t0}
    \langle U(t,\Teff)\rangle \sim (d N \alpha\sigma^2) t^\beta \begin{cases}
        1 & \Teff > T_c\,, \\
        \frac{\rho_c}{\bar{\rho}} & \Teff \leq T_c\,.
    \end{cases}
\end{equation}
For long times, asymptotic analysis reveals that the average potential energy is asymptotically equivalent to a constant for the constant protocol, and to $t^{-d/2}$ for the linear protocol.

At $\Teff = 0$, the internal energy reduces to
\begin{equation}
    \label{eq:internal_energy_T_zero}
    \langle U(t,\Teff=0)\rangle = d N \sigma^2 \gamma(t) \me^{-4 H(t,0)}\begin{cases}
        1 & t < t_c\,, \\
        \frac{\rho_c}{\bar{\rho}} \me^{-2 d H(t,0)} & t \geq t_c\,.
    \end{cases}
\end{equation}

\begin{figure}[htp]
    \centering
    \includegraphics[width=0.75\linewidth]{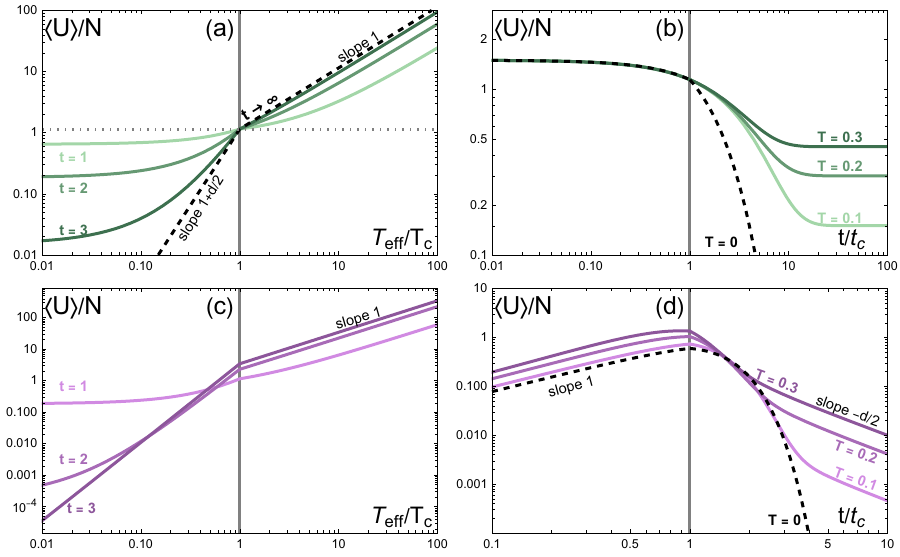}
    \caption{Average potential energy $\langle U\rangle$ for the constant (a-b) and linear (c-d) protocols, as a function of the effective temperature $\Teff$ (a,c) and time $t$ (b,d). The dashed lines indicate the limiting behaviour. Asymptotics are shown by indicating the slope. In (a), the dotted line indicates the value of $\langle U\rangle/N = \alpha d \sigma^2 (\bar{\rho}/\rho_c)^\frac{2}{d} = (d/2) \kb T_c^\infty$, where the energy is fixed, and which is independent of $t$.}
    \label{fig:Ugrid}
\end{figure}

In Fig.~\ref{fig:Ugrid}, we show the average potential energy as a function of effective temperature and time, in panels (a-b) and (c-d) for the constant and linear protocols, respectively.

The heat capacity is then given by the derivative of the internal energy with respect to the effective temperature,
\begin{equation}
    \label{eq:heat_capacity}
    C(t,\Teff) = d N \gamma(t) \pdv{f}{\Teff}\begin{cases}
        1 & \Teff > T_c\,, \\
(1+\frac{d}{2})\frac{\rho_c}{\rho_0} & \Teff\leq T_c\,.
    \end{cases}
\end{equation}

For $\Teff > T_c$, it can be seen that the heat capacity is independent of the effective temperature, regardless of the shape of $\gamma(t)$, since $\partial f/\partial \Teff$ is independent of $\Teff$. For short times, $C$ is asymptotically equivalent to a power law of time, with exponent $1+\beta$, i.e., $C/N\kb \sim t^{1+\beta}$. For long times, however, the heat capacity is once again asymptotically equivalent to a constant for the constant protocol, and to $t^{-d/2}$ for the linear protocol, similar to the asymptotics of the energy.

\begin{figure}[htp]
    \centering
    \includegraphics[width=0.75\linewidth]{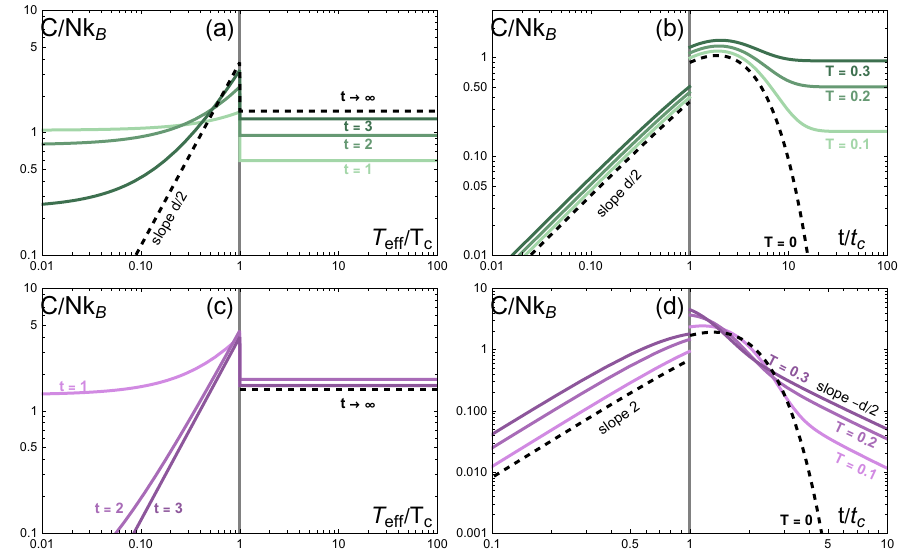}
    \caption{Heat capacity $C/N\kb$ for the constant (a-b) and linear (c-d) protocols, as a function of the effective temperature $\Teff$ (a,c) and time $t$ (b,d). The dashed lines indicate the limiting behaviour. Asymptotics are shown by indicating the slope.}
    \label{fig:cgrid}
\end{figure}

\begin{figure}[htp]
    \centering
    \includegraphics[width=0.5\linewidth]{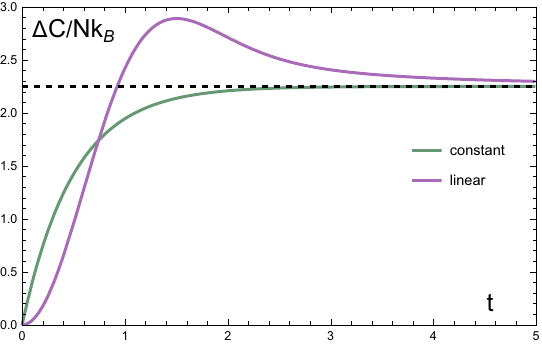}
    \caption{The jump $\Delta C$ in the heat capacity at $\Teff = T_c$ as a function of time for the constant (green) and linear (purple) protocols. The dashed line indicates the asymptotic value $d^2 /4$.}
    \label{fig:delta_C}
\end{figure}

The jump $\Delta C = C(\rho = \rho_c^+) - C(\rho = \rho_c^-)$ in the heat capacity for a fixed time $t$ is given by
\begin{equation}
    \label{eq:heat_capacity_jump}
    \frac{\Delta C}{N \kb} = d^2\gamma(t) \int\limits_0^t \mathrm{d}s\, \me^{-4 H(t,s)}\,,
\end{equation}
which, for the constant and linear protocols, are given respectively by
\begin{equation}
    \label{eq:heat_capacity_jump_specific}
    \begin{split}
        \frac{\Delta C^{(0)}}{N \kb} &= \frac{d^2}{4} \left(1-\me^{-4 \alpha t}\right)\,,\\
        \frac{\Delta C^{(1)}}{N \kb} &= \frac{d^2}{2} \sqrt{\frac{\alpha \pi}{2}} t\, \me^{-2 \alpha t^2} \erfi{(\sqrt{2\alpha}t)} \,,
    \end{split}
\end{equation}
where the superscripts denote the power of $t$ in the protocol ($(0)$ for constant, $(1)$ for linear). Both expressions converge to $d^2/4$ in the long-time limit. The jump in the heat capacity is shown as a function of time in Fig.~\ref{fig:delta_C}. For the constant protocol, $\Delta C$ is an increasing function of time, while for the linear protocol, it reaches a maximum at a finite time before converging to the long-time value.

\begin{figure}
    \centering
    \includegraphics[width=0.75\linewidth]{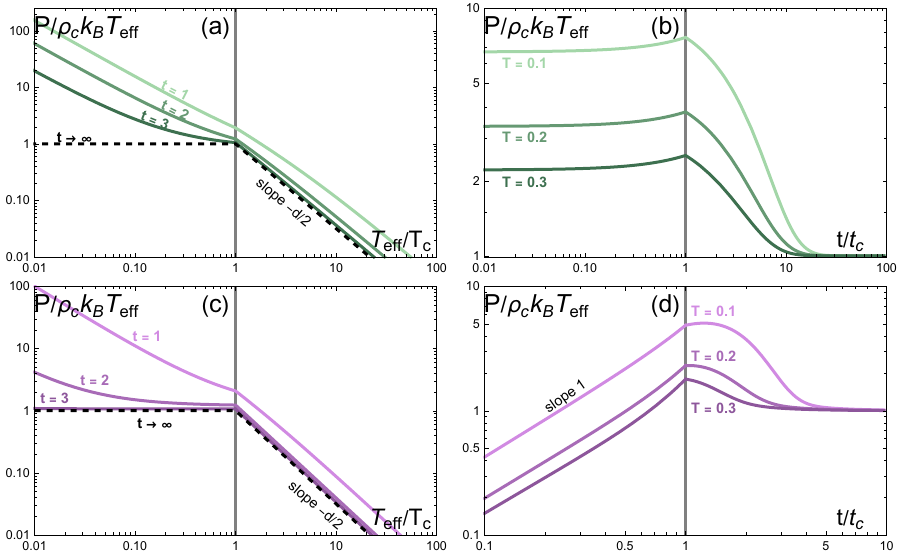}
    \caption{Pressure $P/\rho_c\kb \Teff$ for the constant (a-b) and linear (c-d) protocols, as a function of the effective temperature $\Teff$ (a,c) and time $t$ (b,d). The dashed lines indicate the limiting behaviour. Asymptotics are shown by indicating the slope.}
    \label{fig:Pgrid}
\end{figure}

To compute the bulk pressure, we use a mechanical definition \cite{Solon2015}, which is defined as the average force per unit surface area exerted by particles on a potential $U$ confining the system within a volume $\mathcal{V}$, i.e, 
\begin{equation}
    \label{eq:pressure_mech}
    P_{\rm mech} \equiv \int\limits_0^\infty \mathrm{d}U \rho(U)\,,
\end{equation}
which is identical to the thermodynamic pressure calculated by means of the Helmholtz free energy, $P_{\rm mech} = P_{\rm therm} = P$ \cite{Golestanian_2019,Mahault2020}. Writing the density as a function of the time-dependent potential and integrating, the pressure reads

\begin{equation}
    \label{eq:pressure}
    \begin{split}
    P(t,\Teff) &= 2\gamma(t) f(t,\Teff)\begin{cases}
        \rho_0 & \Teff > T_c \\
        \rho_c & \Teff \leq T_c
    \end{cases}\\
    &= 2N \gamma(t) f(t,\Teff)\begin{cases}
        \mathcal{V}^{-1} & \mathcal{V} > \mathcal{V}_c \\
        \mathcal{V}^{-1}_c & \mathcal{V} \leq \mathcal{V}_c
    \end{cases}
    \end{split}
\end{equation}
where $\mathcal{V} = N/\rho_0 = (N-N_c)/\rho_c$ is the effective volume of the system, with $\mathcal{V}_c = (2 \pi \sigma^2)^\frac{d}{2} (\bar{\rho}/\rho_c)$. For high effective temperature, the pressure goes as $P \sim \Teff^{1-\frac{d}{2}}$, while for low $\Teff$, it behaves asymptotically as a constant that depends on time, i.e., $P \sim 2 \alpha \sigma^2 \rho_c \exp{-4\alpha t^{1+\beta}} t^{\beta}$. For short times, the pressure is asymptotically a power of $t$, following the exponent $\beta$ of the protocol. The pressure as a function of $\Teff$ and $t$ is shown in Fig.~\ref{fig:Pgrid}. 

The second line in eq.~\eqref{eq:pressure} shows that the pressure decouples from the average volume $\mathcal{V}$ when below the critical volume $\mathcal{V}_c$, indicating that the isothermal incompressibility $\kappa^{-1} = -\mathcal{V} (\pdv{P}{\mathcal{V}})_{\Teff}$ diverges at condensation. Eliminating now the effective temperature from the first line of eq.~\eqref{eq:pressure}, we find that the critical point $P_c$ for the pressure isotherms is given by $P_c = \frac{\gamma(t)}{\pi} N^{2/d} \rho_c^{1-\frac{2}{d}}$. Alternatively, for the constant protocol, one can eliminate the time $t$ from the pressure, to find the critical point for the pressure isotemporal line, yielding $P_c = T_c^\infty /\Teff$. While this can formally be done numerically for any type of protocol, analytical progress is unfeasible even for the linear protocol.

Finally, the chemical potential $\mu$ can be defined as the conjugate variable to the particle number $N$, and is given as a derivative of the Helmholtz free energy
\begin{equation}
    \label{eq:chemical_potential}
    \mu = \left(\pdv{\mathcal{F}}{N}\right)_{\Teff,\mathcal{V}}\,.
\end{equation}
After some calculation, we find that $\mu$ is given by
\begin{equation}
    \label{eq:chemical_potential_general}
    \mu = d \gamma(t) f(t,\Teff) \begin{cases}
        -2 \kb \Teff \ln{\left(\frac{\Teff}{T_c}\right)} \left(\frac{\mathcal{G}(t)}{\sigma^2 + 2 \kb \Teff \mathcal{G}(t)}\right) & \Teff > T_c \\
        0 & \Teff \leq T_c
    \end{cases}
\end{equation}
For the constant protocol, the expression for the chemical potential in the non-condensed regime becomes $\mu(\Teff > T_c) = -\frac{d}{2} \kb \Teff \ln{(T/T_c)} (1-\exp{-4 \alpha t})$. In the steady-state regime, this reduces to known results \cite{Mahault2020}.

Equations \eqref{eq:Nc_frac},~\eqref{eq:heat_capacity},~\eqref{eq:pressure} and~\eqref{eq:chemical_potential_general} highlight a remarkable similarity to the thermodynamics of Bose-Einstein condensation, and all thermodynamic observables reduce to the steady-state results obtained in Ref.~\cite{Golestanian_2019} when the protocol is chosen to be constant with $a = k/2$. The dynamical behaviour of our model is moreover akin to the recently derived model for the condensation of a dilute Bose gas subjected to evaporative cooling \cite{Simon2021,Kabelac2022}. Conversely, the discontinuous jump in the heat capacity is a phenomenon that only occurs for dimensions higher than four \cite{Ziff1977}, and it is a consequence of the discrete jump in diffusivity at the condensation threshold $\rho_c$, indicating that the equivalence between our system and BEC is not exact.

\section{Periodic protocols}\label{sec:periodic}
We now turn briefly to the study of periodic protocols, where the amplitude of the harmonic potential is cyclically modulated with a frequency $\omega$. Since it is generally not feasible to analytically solve the integrals involved in eq.~\eqref{eq:f} and associated quantities, we will resort to numerical methods. We discuss only the condensation transition, and refer to future research for the thermodynamic analysis.

\begin{figure}
    \centering
    \includegraphics[width=0.5\linewidth]{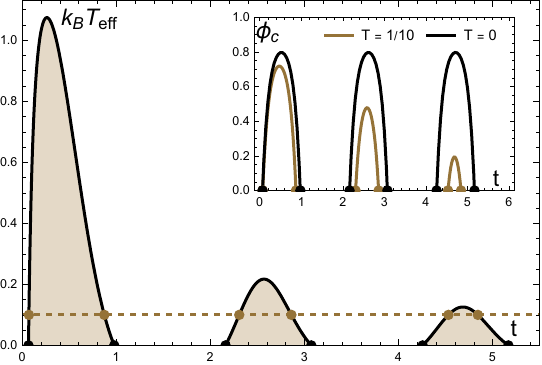}
    \caption{Temperature-time phase diagram for the condensation transition with a periodic protocol. The black and brown dots indicate the condensation times $t_c$ for $\Teff = 0$ and $\Teff = 0.1$ (dashed line), respectively. The inset depicts the number fraction $\phi_c$ for the same fixed temperatures, as a function of time, showing the reentrant behaviour. Parameters are $\sigma = 1$, $\bar{\rho}\rho_c = 2/3$, $\omega = 3$, $\varphi = 0$ and $d=3$.}
    \label{fig:periodic}
\end{figure}

To proceed, we choose the protocol as $\gamma(t) = \cos{(\omega t+\varphi)}$, with $\omega = 2\pi/\tau$ the frequency of the modulation with period $\tau$, and $\varphi$ the phase. We subsequently solve eq.~\eqref{eq:critical_Temp_tc} to find the critical temperature as a function of the critical moment to condensation. The resulting phase diagram is shown in Fig.~\ref{fig:periodic}. As a consequence of the periodic modulation, the condensation is now reentrant as a function of time, for a fixed value of the effective temperature. In the inset, the number fraction is drawn, showing that the number of particles in the condensate is now a non-monotonous function of time.

Note that at $\Teff = 0$, the reentrant behaviour is periodic in time, and the number fraction reaches a global maximum in every condensed phase. Taking the derivative of the number fraction at $\Teff = 0$ with respect to time and setting the result equal to zero leads to the condition 
\begin{equation}
    \label{eq:periodic_condition}
    \pdv{f}{t}\,(t,\Teff = 0) = -4\sigma^2 \pdv{H(t,0)}{t}\me^{-4 H(t,0)} = 0\,.
\end{equation}
Solving $\partial H(t,0)/\partial t = 0$ for $t$ and checking the second derivative shows that the maxima are located at $t_{\rm max} =\tau/4 + j\tau $, $j\in\mathbb{N}$. Inserting $t_{\rm max}$ back into eq~\eqref{eq:Nc_frac} at zero effective temperature shows that the maximum number fraction $\phi_{c,{\rm max}}$ is given by
\begin{equation}
    \label{eq:phi_max}
    \phi_{c,{\rm max}} = 1 - \left(\frac{\rho_c}{\bar{\rho}}\right) \me^{-\frac{2d}{\omega}}\,,
\end{equation}
which is fully determined by the initial system parameters.

\section{Conclusions and outlook}\label{sec:conclusions}

We have studied the dynamics of a non-equilibrium system of particles subjected to both a diffusivity edge and time-dependent harmonic confinement, in arbitrary dimensions. In addition to the temperature-dependent results reported in earlier findings \cite{Golestanian_2019,Mahault2020,Berx_2023}, we have revealed the finite-time emergence of the condensation transition, which can be tuned by both the effective temperature and the time. For constant protocols, and in the steady-state limit, the behaviour and thermodynamics of the active system is equivalent to that of the ideal Bose gas in free space, while the dynamics are closely related to exactly solvable models for evaporative cooling of such gases \cite{Simon2021,Kabelac2022}.

Moreover, we have shown that for any choice of protocol, the time to condensation and the critical temperature are related through equation~\eqref{eq:critical_Temp_tc}. This relation, even at zero effective temperature, emphasises the universal nature of the time-dependence in systems with a diffusivity edge, governed by the interplay of protocol, initial density, and condensation threshold. This universal property enables precise tuning of the time to condensation when these parameters are known. This relation was then extended to a system with periodic modulation of the potential amplitude, leading to a reentrant condensation transition as a function of time.

It would be an interesting avenue of future research to study active systems with a diffusivity edge beyond the mean field approach employed here. Since, e.g., the divergence of the isothermal incompressibility is a result of the  existence of off-diagonal–long-range order (ODLRO) in regular BEC, the current mean field approach is insufficient to capture density correlations leading to this divergence. Finally, extending the current dynamic approach to the BEC-like condensation transition in periodically confined systems would enable experimental modulation of both spatial and temporal periodicity.

\emph{Data availability statement:} All data that support the findings of this study are included within the article (and any supplementary files).

\appendix
\section{Derivation of the time-dependent density profile}\label{app:density}
To find the general time-dependent density profile for the smooth ($\rho < \rho_c$) region, we follow closely the derivation in Ref.~\cite{Fa2016}. Starting from the conservation equation with the single-particle diffusion and mobility coefficients,
\begin{equation}
    \label{eq:conservation}
    \pdv{\rho}{t} = -\div{\vb{J}} = \div\left(D_s \grad\rho - M_s \rho \left(-\grad U\right)\right)\,,
\end{equation}
with the general time-dependent potential $U(\vb{r},t) = \gamma(t) \vb{r}^2 + \beta(t) \vb{r}$, we can rewrite eq.~\eqref{eq:conservation} as follows
\begin{equation}
    \label{eq:conservation_2}
    \pdv{\rho}{t} = D_s \laplacian\rho + M_s \beta(t) \grad\rho + 2\gamma(t) M_s \grad(\rho\vb{r})\,.
\end{equation}
Henceforth, we will suppress the time dependence of $\gamma(t)$ and $\beta(t)$ to ease notation. Taking the $d-$dimensional Fourier transform $\tilde{f} \equiv \mathcal{F}\left\{f(\vb{r})\right\} = \int_{\mathbb{R}^d} \mathrm{d}^d\vb{r}\, f(\vb{r}) \exp{-i\, \vb{k}\vdot\vb{r}}$ of equation~\eqref{eq:conservation_2} yields the following equation for the time derivative of the density
\begin{equation}
    \label{eq:conservation_fourier}
    \pdv{\tilde{\rho}}{t} = -\left(D_s |\vb{k}|^2 + M_s \beta i\vb{k} +2 M_s \gamma \vb{k} \grad\right)\tilde{\rho}\,.
\end{equation}
We seek a solution to equation~\eqref{eq:conservation_fourier} of the form 
\begin{equation}
    \label{eq:ansatz}
    \tilde{\rho}(\vb{k},t) = \prod_{n=1}^\infty\me^{b_n(t) |\vb{k}|^n}\,.
\end{equation}
Substituting \eqref{eq:ansatz} into \eqref{eq:conservation_fourier} gives the following
\begin{equation}
    \label{eq:conservation_fourier_ansatz}
    \sum_{n=1}^\infty\left(b_n'(t) + 2\gamma M_s n b_n(t)\right)|\vb{k}|^n + i M_s \beta \vb{k} + D_s |\vb{k}|^2 = 0\,,
\end{equation}
where a prime indicates the time derivative, such that the coefficients $b_n$ are given by
\begin{equation}
    \label{eq:coefficients}
    \begin{split}
        b_1(t) &= b_{1,0} \me^{-2 M_s H(t,0)} - i M_s \int_0^t\mathrm{d}s\, \beta(s) \me^{-2 M_s H(t,s)}\,, \\
        b_2(t) &= b_{2,0} \me^{-4 M_s H(t,0)} -D_s \int_0^t \me^{-4 M_s H(t,s)}\,,\\
        b_n(t) &= b_{n,0} \me^{-2n M_s H(t,0)} \qquad \mbox{for } n\geq 3\,.
    \end{split}
\end{equation}
Substituting the coefficients back into the solution \eqref{eq:ansatz}
\begin{equation}
    \label{eq:ansatz_filled}
    \begin{split}
    \tilde{\rho}(\vb{k},t) &= \exp{\left(b_{1,0} \me^{-2 M_s H(t,0)} - i M_s \int_0^t\mathrm{d}s\, \beta(s)\me^{-2 M_s H(t,s)}\right) \vb{k}} \\
    &\times \exp{\left(b_{2,0} \me^{-4 M_s H(t,0)} -D_s \int_0^t \me^{-4 M_s H(t,s)}\right) |\vb{k}|^2} \prod_{n=3}^\infty \exp{b_{n,0} \me^{-2n M_s H(t,0)} |\vb{k}|^n}\,,
    \end{split}
\end{equation}
and inserting the Fourier transform of the initial condition~\eqref{eq:initial_condition}, i.e.,
\begin{equation}
    \label{eq:ic_fourier}
    \tilde{\psi}(\vb{k}) = \mathcal{F}\left(\psi(\vb{r})\right) = \me^{\frac{\sigma^2 |\vb{k}|^2}{2}}\,,
\end{equation}
in eq.~\eqref{eq:ansatz_filled} fixes the values of the unknown constants to be $b_{1,0} = b_{3,0} = b_{4,0} = ... = 0$, and $b_{2,0} = -\sigma^2 /2$. Inserting the $b_{n,0}$ back into eq.~\eqref{eq:ansatz_filled} gives
\begin{equation}
    \label{eq:ansatz_filled_2}
    \tilde{\rho}(\vb{k},t) = \exp{-i M_s \vb{k} \int_0^t \mathrm{d}s \beta(s) \me^{-2 M_s H(t,s)}} \exp{-\left(\frac{\sigma^2}{2} \me^{-4 M_s H(t,0)} + D_s \int_0^t\mathrm{d}s\, \me^{-4 M_s H(t,s)}\right)|\vb{k}|^2}\,.
\end{equation}
Finally, taking the inverse Fourier transform $f \equiv \mathcal{F}^{-1}\left\{\tilde{f}(\vb{k})\right\} = (2 \pi)^{-d} \int_{\mathbb{R}^d} \mathrm{d}^d\vb{k}\, f(\vb{k}) \exp{i\, \vb{k}\vdot\vb{r}}$ of the solution~\eqref{eq:ansatz_filled_2}, we arrive at
\begin{equation}
    \label{eq:final_solution}
    \rho(\vb{r},t) = \left(2 \pi f(t,T)\right)^{-\frac{d}{2}} \exp{-\frac{\left(\vb{r} - g(t)\right)^2}{2 f(t,T)}}\,,
\end{equation}
with $f(t,T) = \sigma^2 \me^{-4M_s H(t,0)} + 2M_s\kb \Teff\int_0^t \mathrm{d}\tau\, \me^{-4 M_s H(t,\tau)}$, which is equation~\eqref{eq:f} in the main text, and $g(t) = M_s \int_0^t \mathrm{d}s\, \beta(s) \me^{-2 M_s H(t,s)}$. 

\bibliographystyle{apsrev4-2}
\bibliography{biblio.bib}
\end{document}